\documentclass[twocolumn]{aastex631}

\usepackage{mathtools}

\shorttitle{SN$\nu$NSI}
\shortauthors{Huang et al.}

\graphicspath{{./}{figures/}}
\begin{document}

\title{Supernova Preshock Neutronization Burst as a Probe of Non-Standard Neutrino Interactions}

\correspondingauthor{Lie-Wen Chen}
\email{lwchen@sjtu.edu.cn}

\author[0000-0003-1842-8657]{Xu-Run Huang}
\affiliation{School of Physics and Astronomy, Shanghai Key Laboratory for
Particle Physics and Cosmology, and Key Laboratory for Particle Astrophysics and Cosmology (MOE),
Shanghai Jiao Tong University, Shanghai 200240, China}

\author[0000-0001-6773-7830]{Shuai Zha}
\affiliation{Tsung-Dao Lee Institute, Shanghai Jiao Tong University, Shanghai 200240, China}

\author[0000-0002-7444-0629]{Lie-Wen Chen}
\affiliation{School of Physics and Astronomy, Shanghai Key Laboratory for
Particle Physics and Cosmology, and Key Laboratory for Particle Astrophysics and Cosmology (MOE),
Shanghai Jiao Tong University, Shanghai 200240, China}

\begin{abstract}
Core-collapse supernova (CCSN) provides a unique astrophysical site for studying neutrino-matter interactions. Prior to the shock-breakout neutrino burst during the collapse of the iron core, a preshock $\nu_e$ burst arises from the electron capture of nuclei and it is sensitive to the low-energy coherent elastic neutrino-nucleus scattering (CE$\nu$NS) which dominates the neutrino opacity. Since the CE$\nu$NS depends strongly on the non-standard neutrino interactions (NSI) which are completely beyond the standard model and yet to be determined, the detection of the preshock burst thus provides a clean way to extract the NSI information. Within the spherically symmetric general-relativistic hydrodynamic simulation for the CCSN, we investigate the NSI effects on the preshock burst. We find that the NSI can maximally enhance the peak luminosity of the preshock burst almost by a factor of three, reaching a value comparable with that of the shock-breakout burst. Future detection of the preshock burst will have critical implications on astrophysics, neutrino physics and physics beyond the standard model.
\end{abstract}

\section{Introduction}
Neutrinos interact feebly with ordinary matter \citep{1999RvMPS..71..140W}. Nonetheless, they play a critical role in a core-collapse supernova~(CCSN), which marks the death of massive stars with mass $\gtrsim8~M_\odot$ and leaves behind a compact remnant \cite[see][for reviews]{Woosley:2002zz,2012ARNPS..62..407J,2021Natur.589...29B}. In a CCSN, most ($\sim 99 \%$) of the released gravitational potential energy ($\sim 10^{53}~\rm{erg}$) of the progenitor star is ultimately liberated through neutrino emission within a $\sim 10$~s burst. Neutrinos from CCSNe can thus  carry invaluable information on both CCSN and neutrino physics~\citep{2003RvMP...75.1011K}.

Meanwhile, the discovery of neutrino oscillations~\citep{1998PhRvL..81.1562F,2002PhRvL..89a1301A} indicates neutrinos are massive and lepton flavors are mixed, providing solid experimental evidence of physics beyond the Standard Model~(SM).
Current and upcoming neutrino experiments
can measure subdominant neutrino oscillation effects that are expected to give information on the yet-unknown neutrino parameters and the non-standard interactions~(NSI) between neutrinos and matter~\citep{1978PhRvD..17.2369W,2013RPPh...76d4201O,2018FrP.....6...10T,Dev:2019anc}.
Note that the NSI are completely originated from new physics beyond the SM and not an expected consequence of existing theories or
neutrino oscillations. The NSI can modify the production, propagation and detection of neutrinos and thus may crucially affect the interpretation of the relevant experimental data.
While the oscillation experiments
can put important constraints on the NSI parameters, non-oscillation data (e.g.,
from neutrino scattering experiments) is needed to break the possible degeneracy of the neutrino parameters allowed by oscillation data alone~\citep{2017JHEP...04..116C}.
Indeed, the deep inelastic neutrino scattering experiments (e.g., CHARM~\citep{1986PhLB..180..303D} and NuTeV~\citep{2002PhRvL..88i1802Z}) can help to break degeneracy but the constraints apply only if the NSI are generated by mediators not much lighter than the electroweak scale.
For light mediators~\citep{Denton:2018xmq},
the degeneracy can only be broken through combining with results on coherent elastic neutrino-nucleus scattering~(CE$\nu$NS), which was predicted in the 1970's~\citep{1974PhRvD...9.1389F} but observed only recently by the COHERENT Collaboration~\citep{Akimov:2017ade,2021PhRvL.126a2002A}.

Indeed, a global fit to neutrino oscillation and CE$\nu$NS data indicates that the degeneracy of neutrino parameters is significantly disfavored for a wider range of NSI models~\citep{2018FrP.....6...10T,2018JHEP...08..180E,2020JHEP...02..023C}.
Although significant progress
has been made on constraining the NSI parameters by analyzing data on neutrino oscillations, deep inelastic neutrino scattering and CE$\nu$NS, some NSI parameters are still not well constrained. In particular, the vectorlike quark-$\nu_e$ neutral
current~(NC) couplings, $\varepsilon_{ee}^{uV}$ and $\varepsilon_{ee}^{dV}$, are the least experimentally
constrained~\citep{2018FrP.....6...10T,2018JHEP...08..180E,2020JHEP...02..023C}, preserving parameter space large enough
for causing sizeable modifications on the CE$\nu$NS
cross sections.
Compared to the case of charged current (CC) NSI, it is a much more daunting task to
constrain NC NSI due to the experimental and theoretical difficulties.
%in measuring neutrino NC interactions, and to the technical difficulties in computing neutrino-nucleus interactions.
Because of the frequent CE$\nu$NS in a CCSN, the CCSN can thus provide an ideal site for constraining the NC NSI parameters.
Recently, \cite{2021PhRvD.103h3002S} estimate the NSI effects on neutrino-nucleon scattering in the post-bounce SN core within the diffusion time criterion wherein the neutrinos cannot be trapped for too long.
In this work,
we show that the neutrino burst from the preshock neutronization in a CCSN can be used as a novel and clean probe of the NC NSI parameters $\varepsilon_{ee}^{uV}$ and $\varepsilon_{ee}^{dV}$.

\section{Preshock Neutronization Burst}

Modern CCSN models~\citep{2018JPhG...45j4001O} have commonly
predicted the existence of the so-called neutronization neutrino burst with a peak luminosity $\sim 4\times 10^{53}$~erg$\cdot$s${^{-1}}$, which emerges during the first $\sim 25$~ms after the core bounce
as a result of sudden breakout of a flood of neutrinos freshly produced in shock-heated matter (and some
$\nu_e$ produced previously that have diffused to the neutrinosphere) when the
bounce shock penetrates the neutrinosphere and reaches the neutrino-transparent regime at sufficiently low densities.
This shock-breakout burst mainly comprises $\nu_e$ from electron captures on free protons in the shock-heated matter.

Prior to the shock-breakout burst,
there exists a smaller burst due to $\nu_e$ produced from the preshock neutronization of the collapsing core~\citep{2003NuPhA.719..144L,2005PhRvD..71f3003K,2016ApJ...817..182W,2018JPhG...45j4001O}.
This preshock burst emerges as a result of the competition between the $\nu_e$ emission due to electron captures on nuclei during the early neutronization stage of core collapse and the $\nu_e$ trapping due to the opacity enhancement as the density and temperature of the core
increase. Although the preshock burst is weaker than the shock-breakout burst, it generally has weaker model dependence in the CCSN simulations since it only involves the relatively simpler dynamics in the early stage of the CCSN.
In particular, the preshock burst is expected to strongly depend on the CE$\nu$NS cross sections which essentially control the neutrino opacity in the preshock stage~\citep{1997PhRvD..56.7529B}, and thus to provide a clean probe of the NC NSI parameters $\varepsilon_{ee}^{uV}$ and $\varepsilon_{ee}^{dV}$. It should be noted that the NC interactions change the neutrino
opacity without directly changing the neutrino production rate which is mainly determined by the CC electron capture processes~\citep{2016ApJ...816...44S,2021RPPh...84f6301L}.

\section{NSI effects on neutrino-nucleus scattering}
Following the spirit of effective four-fermion couplings in low-energy weak interactions, the NC NSI Lagrangian can be typically formulated as~\citep{2013RPPh...76d4201O,2018FrP.....6...10T,Dev:2019anc}
\begin{equation}
\label{eq:Lagrangian_NSI}
    \mathcal{L}_{\mathrm{NSI}} = -2\sqrt{2} G_F \varepsilon_{\alpha \beta}^{f X} ( \bar{\nu}_\alpha \gamma^\mu P_L \nu_\beta ) (\bar{f} \gamma_\mu P_X f) ,
\end{equation}
where $G_F$ is the Fermi constant, $\varepsilon_{\alpha \beta}^{f X}$ denote the NSI parameters with $\varepsilon_{\alpha \beta}^{f X}\sim 1$ corresponding to a NSI strength comparable to that of SM weak interactions, $\alpha, \beta \in \{e, \mu, \tau\}$ represent neutrino flavors, $f \in \{e, u, d \}$ is the matter field, and $P_{X}$ with $X = L(R)$ represents the left(right) chirality projection operator. The NSI parameters are flavor diagonal for $\alpha = \beta$, while the lepton flavor is violated and the NSI become flavor-changing for $\alpha \neq \beta$. Here we mainly focus on the flavor-diagonal NC vectorial NSI couplings of $\nu_e$ to the light quarks, i.e.,
\begin{equation}
\label{eq:NSIpar}
\varepsilon^{fV}_{ee} = \varepsilon_{ee}^{fL} + \varepsilon_{ee}^{fR}, f \in \{u, d\} ,
\end{equation}
since they have relatively larger parameter space with $\varepsilon^{uV}_{ee} (\varepsilon^{dV}_{ee}) \in [0.0, 0.5]$ while the amplitude of other NSI parameters has been tightly constrained to be $\lesssim 0.1$~\citep{2018FrP.....6...10T,2018JHEP...08..180E,2020JHEP...02..023C}. Note the SNO results~\citep{2008PhRvL.101k1301A} agree well with the prediction of standard solar model, suggesting a small NSI axial interactions and thus $\varepsilon_{ee}^{fL} \approx \varepsilon_{ee}^{fR}$.
With $\varepsilon^{u(d)} \equiv \varepsilon^{u(d)V}_{ee}$, the effective NSI couplings to nucleons can thus be obtained as
\begin{equation}
\label{eq:NSItoNucleon}
\varepsilon^p = 2 \varepsilon^u + \varepsilon^d , \ \ \  \varepsilon^n = \varepsilon^u + 2 \varepsilon^d.
\end{equation}

For neutrino-matter interactions, we use here the neutrino interaction library \emph{NuLib}~\citep{2015ApJS..219...24O}. In order to investigate the effects of the NC NSI parameters $\varepsilon^{u}$ and $\varepsilon^{d}$, we modify the cross sections of the following iso-energetic reactions,
$\nu_e + \alpha \longleftrightarrow \nu_e + \alpha, \nu_e + p \longleftrightarrow \nu_e + p, \nu_e + n \longleftrightarrow \nu_e + n, \nu_e + \prescript{A}{Z}{X} \longleftrightarrow \nu_e + \prescript{A}{Z}{X}$,
and the corresponding reactions induced by $\bar{\nu}_e$.
For (anti-)neutrino-nucleus scattering, the cross section includes three corrections~\citep{2006NuPhA.777..356B}: the ion-ion correlation function $\langle \mathcal{S}_{\mathrm{ion}} \rangle$, the form factor term $\mathcal{C}_{\mathrm{FF}}$ and the electron polarization correction $\mathcal{C}_{\mathrm{LOS}}$. The expressions of the three corrections keep unchanged since they are irrelevant to the
NC NSI parameters $\varepsilon^{u}$ and $\varepsilon^{d}$.
For simplicity, we neglect the weak magnetism corrections for anti-neutrinos~\citep{2002PhRvD..65d3001H} since here we mainly focus on the neutronization burst in the early stage of CCSN, which mainly involves $\nu_e$.
In such a case, the cross section modification is rather straightforward, namely, we only need to replace the NC vector couplings $g_V^p = 1/2 - 2 \sin^2 \theta_W$ and $g_V^n = - 1/2$ in the SM, respectively, by $g_V^{p \prime}$ and $g_V^{n \prime}$ as
\begin{equation}
\label{eq:VectCoup}
g_V^{p \prime} = g_V^p + \varepsilon^p , \ \ \  g_V^{n \prime} = g_V^n + \varepsilon^n .
\end{equation}
Correspondingly, the cross section expression is modified by replacing the weak charge of nucleus $\mathcal{Q}_W = -2 (Z g_V^p + N g_V^n)$ by $\mathcal{Q}_W^{\prime}$ as
\begin{equation}
\label{eq:WeakCharge}
\mathcal{Q}_W^{\prime} = \mathcal{Q}_W + \mathcal{Q}_W^{\mathrm{NSI}}, \ \ \ \mathcal{Q}_W^{\mathrm{NSI}} \equiv -2 (Z \varepsilon^p + N \varepsilon^n).
\end{equation}

\begin{figure}[t!]
 \centering
 \includegraphics[width=0.95\columnwidth]{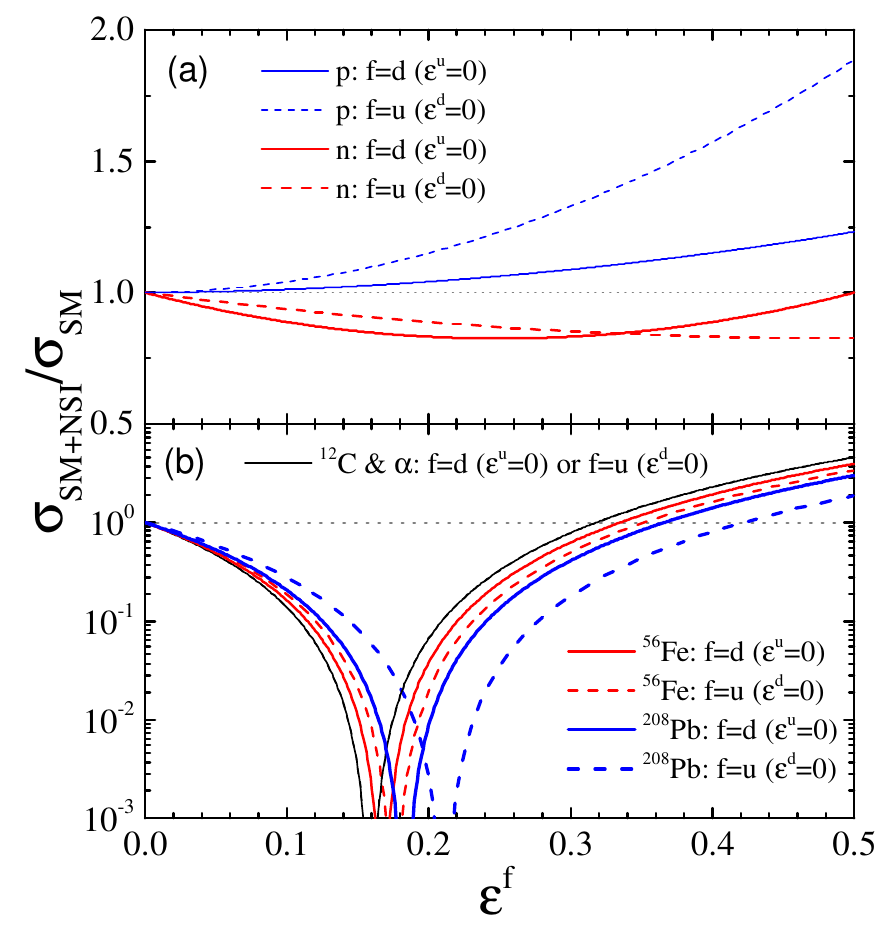}
 \caption{Neutrino-nucleon (a) and neutrino-nucleus (b) scattering cross sections divided by their SM model values as functions of the NSI parameter $\varepsilon^d$ ($\varepsilon^u = 0$) or $\varepsilon^u$ ($\varepsilon^d = 0$).}
\label{fig:CroSec}
\end{figure}

The ratio of the neutrino-nucleus cross sections with and without NSI can be expressed as $\sigma_{\mathrm{SM + NSI}} / \sigma_{\mathrm{SM}} = \mathcal{Q}_W^{\prime 2} / \mathcal{Q}_W^2$~\citep{2006NuPhA.777..356B} if we neglect the corrections from $\mathcal{C}_{\mathrm{FF}}$ and $\mathcal{C}_{\mathrm{LOS}}$.
To examine the NSI effects on neutrino-nucleus scattering, we plot in Fig.~\ref{fig:CroSec} the ratio $\sigma_{\mathrm{SM + NSI}} / \sigma_{\mathrm{SM}}$ as a function of $\varepsilon^d$ ($\varepsilon^u = 0$) or $\varepsilon^u$ ($\varepsilon^d = 0$) for several typical nuclei, i.e., $\alpha$, $^{12}$C, $^{56}$Fe and $^{208}$Pb, as well as protons (p) and neutrons (n).
One sees that the neutrino-nucleus cross sections can be drastically suppressed and even vanish around a certain value of $\varepsilon^d$ ($\varepsilon^u$) depending on the isospin of the nucleus. This is due to the fact that the effective weak charge $\mathcal{Q}_W^{\prime}$ may vanish for a certain value of $\varepsilon^d$ ($\varepsilon^u$) satisfying the relation $Y_p (g^p_V + 2\varepsilon^u + \varepsilon^d) + (1 - Y_p) (g^n_V + \varepsilon^u + 2\varepsilon^d) = 0$, where $Y_p = Z/A$ is the proton fraction of the nucleus. One can easily find $\mathcal{Q}_W^{\prime} = 0$ when $\varepsilon^u + \varepsilon^d = 0.159$ for nuclei with $N=Z$ (e.g., $\alpha$, $^{12}$C), and for more neutron-rich nuclei (e.g., $^{56}$Fe, $^{208}$Pb) with smaller $Y_p$, $\mathcal{Q}_W^{\prime} = 0$ generally leads to larger $\varepsilon^d$ ($\varepsilon^u$), as shown in Fig.~\ref{fig:CroSec}.
On the other hand, the neutrino-p(n) cross section exhibits relatively weak sensitivity to $\varepsilon^d$ or $\varepsilon^u$.
These features will lead to a number of interesting consequences on the neutrino burst in CCSN.

%-------------------------------------------------------------------------------------------------------------------
%-------------------------------------------------------------------------------------------------------------------

\section{NSI effects on neutrino burst}
\begin{figure}[t!]
 \centering
 \includegraphics[width=0.95\columnwidth]{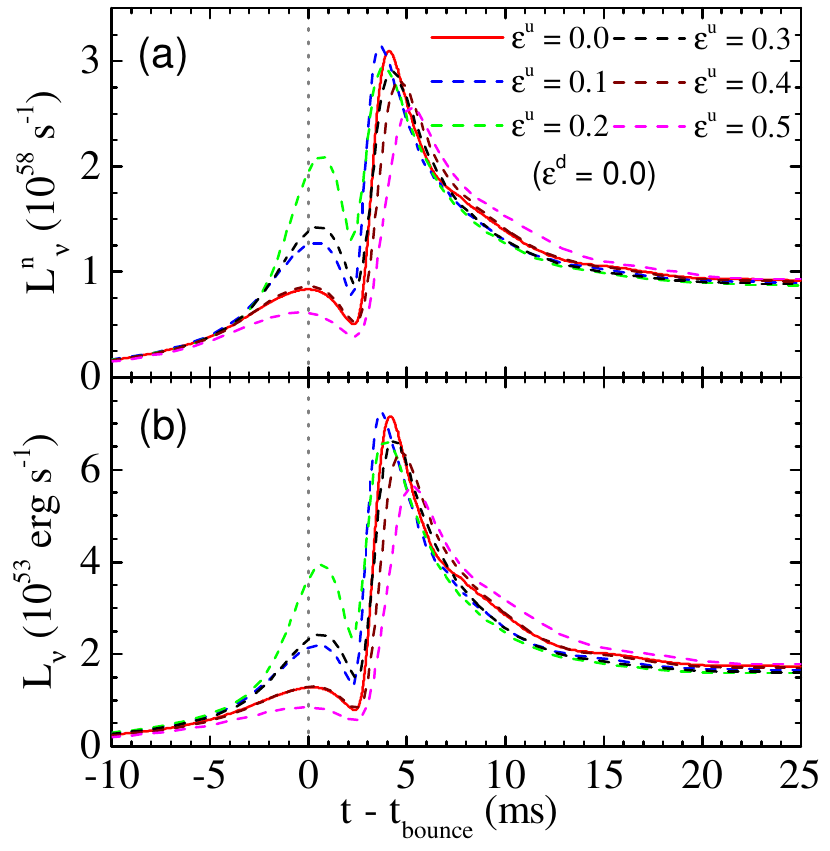}
 \caption{Time evolution of the total neutrino number~(a) and energy~(b) luminosities for the stellar collapse of a $15 \ M_\odot $ solar-metallicity progenitor star using the SFHo EOS with various $\varepsilon^{\rm{u}}$ values ($\varepsilon^{\rm{d}}=0$).}
\label{fig:numLumNSI}
\end{figure}

SN core collapse and bounce are simulated using the spherically-symmetric general-relativistic hydrodynamic code \emph{GR1D}~\citep{2010CQGra..27k4103O,2015ApJS..219...24O}. As a default of the CCSN simulation, we adopt the $15 \ M_\odot $ solar-metallicity progenitor star (s15s7b2) from \cite{1995ApJS..101..181W}, and the SFHo equation of state (EOS) from \cite{2013ApJ...774...17S} is used to describe the physics of stellar matter.
Fig.~\ref{fig:numLumNSI} shows the time evolution of all-flavor neutrino number and energy luminosities in the initial two stages of CCSN, i.e., the infall phase and neutronization burst, with $\varepsilon^u = 0, 0.1, 0.2, 0.3, 0.4, 0.5$ ($\varepsilon^d = 0$). The later two stages of accretion phase and Kelvin-Helmholtz cooling phase are not shown for simplicity since our main focus is the preshock burst.

%-------------------------------------------------------------------------------------------------------------------
%-------------------------------------------------------------------------------------------------------------------

For all the $\varepsilon^u$ values considered here, it is clearly seen from Fig.~\ref{fig:numLumNSI} that the luminosity displays two peaks, i.e., the smaller one around the bounce and the larger one after the bounce, respectively corresponding to the preshock burst and the shock-breakout burst.
In particular, we note (although not shown here) that the preshock burst essentially consists of only $\nu_e$, and the shock-breakout burst (around the peak) is also dominated by $\nu_e$ with $\lesssim 15\%$ heavy-flavor (anti-)neutrinos and tiny ($\lesssim 1\%$) $\bar{\nu}_e$. In addition, the average $\nu_e$ energy of the preshock burst is $\sim 10$~MeV. For the shock-breakout burst, the average energy is $\sim 14$~MeV for $\nu_e$, $\sim 15$~MeV for heavy-flavor (anti-)neutrinos and $\sim 10$~MeV for $\bar{\nu}_e$.
These general features have been also observed in various modern CCSN simulations~\citep{2018JPhG...45j4001O}.

The most interesting feature illustrated in Fig.~\ref{fig:numLumNSI} is the NSI effects on the two bursts, i.e., while the variation of the peak luminosity for the shock-breakout burst with $\varepsilon^u$ is a little complicated and relatively weak ($\lesssim 10\%$), the corresponding variation for the preshock burst is rather straightforward and very drastic.
For the latter, the peak luminosity first increases with $\varepsilon^u$ varying from $0$ to $0.2$, and then decreases as $\varepsilon^u$ changes from $0.2$ to $0.5$. Such a variation is mainly due to the NSI effects on the neutrino-nucleus scattering. As shown in Fig.~\ref{fig:CroSec}, increasing $\varepsilon^u$ from $0$ to $\sim 0.2$ will reduce drastically the neutrino-nucleus cross section and even make it vanish at $\varepsilon^u \sim 0.2$, and the cross section enhances again as $\varepsilon^u$ increases from $\sim 0.2$ to $0.5$.
During the early neutronization stage of CCSN, the $\nu_e$, $e^-$ and nuclei are dominant and the CE$\nu$NS decisively controls the neutrino opacity~\citep{1997PhRvD..56.7529B}. The suppression on neutrino-nucleus scattering will increase neutrino's mean free path and thus enhance the neutrino emission.
Quantitatively, it is remarkable to see from Fig.~\ref{fig:numLumNSI} that the peak number (energy) luminosity of the preshock burst can reach to $\sim 2.1\times 10^{58}$~s${^{-1}}$ ($\sim 3.9\times 10^{53}$~erg$\cdot$s${^{-1}}$) for $\epsilon^u = 0.2$, which is significantly larger than and almost three times of the corresponding value of without NSI [i.e., $\sim 0.86\times 10^{58}$~s${^{-1}}$ ($\sim 1.3\times 10^{53}$~erg$\cdot$s${^{-1}}$) for $\epsilon^u = 0$], and it even becomes comparable to the corresponding result of the shock-breakout burst [i.e., $\sim 2.5\times 10^{58}$~s${^{-1}}$ ($\sim 6.0\times 10^{53}$~erg$\cdot$s${^{-1}}$)].

It is interesting to see that the peak luminosity of the shock-breakout burst does not much depend on the NSI, and this is understandable since the shock-breakout burst neutrinos are mainly produced through electron captures on free protons in the shock-heated matter and escape in the neutrino-transparent regime at sufficiently low densities where the neutrino-nucleus scattering is less important.
The neutrino-nucleon scattering in the shock-heated matter may give rise to opacity and thus influence the neutrino emission of the shock-breakout burst, but the NSI effects are relatively weak as shown in Fig.~\ref{fig:CroSec}~(a).
Moreover, the preshock burst may also slightly influence the shock-breakout burst since the former affects the $\nu_e$'s distribution behind the neutrinosphere.
Furthermore, modern CCSN simulations~\citep{2018JPhG...45j4001O} indicate some sensitivity of the shock-breakout burst height and shape to the details of the neutrino transport, while the preshock burst is relatively robust due to the much simpler dynamics involved. Therefore, our results suggest that the preshock burst of CCSN should be a clean probe of the NSI.

%-------------------------------------------------------------------------------------------------------------------
\begin{figure}[t!]
 \centering
 \includegraphics[width=0.95\columnwidth]{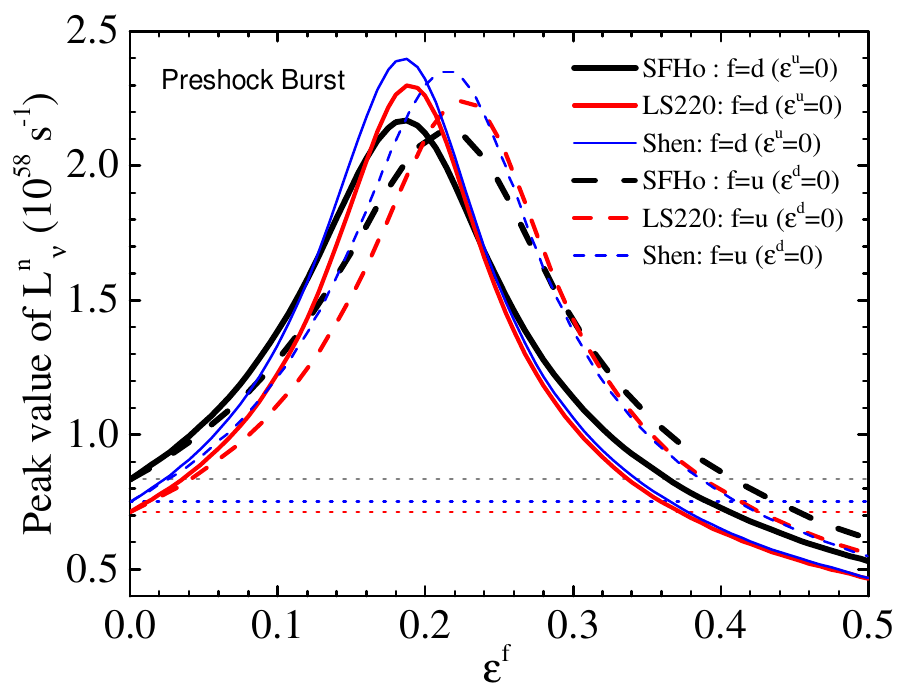}
 \caption{The total number luminosity peak value of the preshock burst versus the NSI parameter $\varepsilon^f$ for the stellar collapse of the $15 \ M_\odot $ solar-metallicity progenitor star with three different EOSs, i.e., SFHo, LS220 and Shen. The dotted lines show the SM values for each EOS.}
\label{fig:DependenceEOS}
\end{figure}
%-------------------------------------------------------------------------------------------------------------------

To examine the robustness of the preshock burst as a probe of $\varepsilon^f$, we show in Fig.~\ref{fig:DependenceEOS} the peak number luminosity of the preshock burst as a function of $\varepsilon^d$ ($\varepsilon^u = 0$) and $\varepsilon^u$ ($\varepsilon^d = 0$) using three different EOSs, i.e., the default SFHo EOS, the LS220 EOS from \cite{Lattimer:1991nc} with nuclear matter incompressibility $K_{0} = 220$~MeV and Shen EOS from \cite{2011ApJS..197...20S}. One sees that the difference of the peak number luminosity from the three EOSs is relatively small ($\sim 10\%$).
The weak EOS dependence is mainly due to the small difference of low density ($\lesssim 10^{12}$~g/cm$^3$) stellar matter EOS for the three EOSs since the preshock burst mainly involves stellar matter with density up to the neutrino trapping value ($\sim 10^{12}$~g/cm$^3$).
Indeed, using the Lattimer and Swesty EOSs~\citep{Lattimer:1991nc} with $K_{0} = 180$~MeV and $375$~MeV which give very different EOS around and above nuclear density ($\gtrsim 10^{14}$~g/cm$^3$) but same low density EOS, we find the resulting peak number luminosities are almost the same as that with $K_{0} = 220$~MeV.
In addition, one sees from Fig.~\ref{fig:DependenceEOS} that the $\varepsilon^f$ maximizing the peak number luminosity is larger than that minimizing the $\nu_e$-$^{56}$Fe cross section as shown in Fig.~\ref{fig:CroSec}, and this is mainly because the $^{56}$Fe nuclei in the collapsing core are transformed to more neutron-rich nuclei due to electron captures and thus larger $\varepsilon^f$ is needed to minimize the $\nu_e$-nucleus cross sections as discussed previously.

We also note the preshock burst only weakly depends on the progenitor mass, consistent with the earlier findings~\citep{2003PhRvD..68k3009T,2005PhRvD..71f3003K,2016ApJ...817..182W}. Nevertheless, the progenitor property can be constrained with multimessenger signals~\citep{2020ApJ...899..153M,2013ApJ...762..126O,2020ApJ...898..139W,2021arXiv210110624S,2021arXiv210201118B} once the source is detected.
Moreover, the more realistic three-dimensional~(3D) simulations~\citep{2021MNRAS.500..696N} give very similar predictions on the neutronization burst during the early stage of CCSN as the one-dimensional~(1D) simulations adopted here, further justifying the robustness of the preshock burst as a probe of $\varepsilon^f$.
In addition, the non-standard neutrino self interactions (NSSI) are not considered here. Although the NSSI may significantly modify the neutrino flavor transformation and thus influence the neutrino spectra~\citep{2018PhRvD..97d3011D,2018PhRvD..97j3018Y,2020JHEP...01..179L}, they are not expected to cause sizeable modification on our results unless the NSI neutrino-neutrino coupling $g_{\nu}$ can be significantly larger than the NSI neutrino-quark coupling $g_q$  (e.g., $g_{\nu} \gtrsim 90g_q$). This is because the NSSI only have minor impact on the neutrino opacity due to the small $\nu_e$ fraction and cross section compared to those of nuclei in the early collapsing core. It will be interesting to see the NSSI effects on the preshock burst when the $g_{\nu}$  is extremely large (e.g., $g_{\nu} \gtrsim 90g_q$).

In Fig.~\ref{fig:DependenceEOS}, we consider only two extreme cases by independently varying $\varepsilon^d$ ($\varepsilon^u = 0$) or $\varepsilon^u$ ($\varepsilon^d = 0$), and the results with simultaneous variation of $\varepsilon^d$ and $\varepsilon^u$ should be between the corresponding results of the two extreme cases.
Moreover,
due to the quadratic dependence of the CE$\nu$NS cross section on the weak charge, there inevitably exists $\varepsilon^f$ degeneracy for a fixed peak luminosity of the preshock burst. In particular, Fig.~\ref{fig:DependenceEOS} displays degeneracy for  $\varepsilon^f = 0$ and $\varepsilon^f \sim 0.4$.
The combined analysis of neutrino oscillation and CE$\nu$NS experiments perhaps can break the degeneracy. As pointed out in \citep{2018FrP.....6...10T}, $\varepsilon^d \simeq 0.3$ is more favored than $\varepsilon^d = 0$ at a level of $2 \sigma$ in analyses of solar neutrino experiments.
Recently, the COHERENT collaboration report their new measurement of CE$\nu$NS on Argon, excluding the parameter region around $\varepsilon^f \sim 0.2$ with $90 \%$ C.L.~\citep{2021PhRvL.126a2002A}.
Nevertheless, the peak luminosity of the preshock burst still keeps great sensitivity to the NSI in the remaining parameter space.

It is instructive to have a discussion on the experimental detection of the preshock burst.
Although
the neutrino oscillation should not lead to magnificent modifications to the core collapse dynamics~\citep{2011PhRvL.107o1101C,2012PhRvD..85f5008D,2020PhRvD.102h1301S}, it will largely distort the $\nu_e$ emission pattern in terrestrial detectors. Hence, it is better to use all-flavor detection to depict the temporal structure of the preshock burst.
Recently, \cite{2020PhRvL.124n1802R}
shows the feasibility of detecting neutrino number luminosity from a failed CCSN using large-scale DM detectors, from which we note the detection of the preshock burst is possible if a source is located within $\sim 1 $~kpc. Luckily, such pre-supernova stars are somewhat not too rare in our galaxy, and a list of $31$ candidates within $1$~kpc, including the famous Betelgeuse, is rendered in \cite{2020ApJ...899..153M}.
In addition,
the $\varepsilon^f$ reduces the detection rate of the detectors made of nuclei but has no effects on the neutrino-electron cross sections and even enhances the neutrino-p cross sections, and therefore the detectors made of protons or electrons should be ideal choice.
As an example,
we estimate the detection potential of $\varepsilon^u$ by the Hyper-Kamiokande~\citep{2018arXiv180504163H} via the electron scattering channel.
Using the \emph{sntools}~\citep{2021JOSS....6.2877M} code to simulate the detector response for the preshock burst from a 1 kpc CCSN,
we find the event count per 1 ms can reach $\mathcal{O}(10^2)$ around the preshock burst peak.
By assuming $\chi^2 = \sum (N_{\rm SM} - N_{\rm NSI})^2 /\sigma_{\rm stat.}^2$,
we find the discovery region of $\varepsilon^u$ with $3\sigma$ is
$[0.015, 0.388]\bigoplus[0.415, 0.5]$ for no neutrino oscillation and $[0.033(0.023), 0.373(0.380)]\bigoplus[0.438(0.424), 0.5]$ for the oscillation scenario with normal (inverted) neutrino-mass ordering.
Furthermore, it is important to note that
the flavor-blind measurement via the elastic neutrino-p scattering, e.g., in JUNO~\citep{2016JPhG...43c0401A}, can avoid the influence of oscillation and even break the degeneracy at $\varepsilon^f \sim 0.4$ due to the NSI enhancement of neutrino-p cross sections as shown in Fig.~\ref{fig:CroSec}~(a). Such detection configuration of JUNO is yet to be added in \emph{sntools}.

Finally, we note that the enhancement of neutrino emission in the preshock burst can reduce the central electron fraction $Y_e$ of the CCSN, e.g., the central $Y_e$ after the bounce is reduced from $0.281$ to $0.247$ as $\varepsilon^u$ varies from $0$ to $0.2$. This reduction of $Y_e$ may influence the later neutrino flavor evolution, explosion dynamics and nucleosynthesis~\citep{2019PrPNP.107..109K,2021RvMP...93a5002C} of the CCSN. Reliable predictions on these topics are beyond the 1D simulations, and it will be extremely interesting to explore them within the more realistic 3D simulations~\citep{2021MNRAS.500..696N}. In addition, it is worth noting that the detection of the preshock burst may provide a clean way to extract neutrino oscillation information and determine the neutrino-mass hierarchies~\citep{2003PhRvD..68k3009T,2005PhRvD..71f3003K,2016ApJ...817..182W}.

\section{Conclusion}
We have demonstrated that the preshock neutrino burst in CCSN can serve as a clean probe of the largely unknown NSI parameters $\varepsilon_{ee}^{uV}$ and $\varepsilon_{ee}^{dV}$.
In particular, our results indicate that the NSI can enhance the peak luminosity of the preshock burst almost by a factor of three and make the luminosity comparable to that of the shock-breakout burst, which will have critical implications on the explosion dynamics of CCSN.
Future detection of
the preshock burst will open a new window to extract information on the CCSN, the NSI, the neutrino oscillation, and the neutrino-mass hierarchies.

\section*{Acknowledgements}
The authors would like to thank Jianglai Liu, Chuanle Sun and Donglian Xu for useful discussions.
This work was supported by National SKA Program of China No. 2020SKA0120300 and the National Natural Science Foundation of China under Grant No. 11625521.

\software{GR1D~\citep{2010CQGra..27k4103O,2015ApJS..219...24O}, NuLib~\citep{2015ApJS..219...24O}, sntools~\citep{2021JOSS....6.2877M}.}


\bibliography{SNvNSI}{}
\bibliographystyle{aasjournal}



\end{document}